\documentstyle{l-aa}      

\newcommand{\ebv}{$E(B-V)$}
\newcommand{\phm}{\phantom{$-$}}

\begin{document}
\input{psfig.tex}
\thesaurus{ 03     
        (08.19.4	 
	 08.19.5 ) 
                   }
\title{Revised photometry and color distribution
 of Type Ia
supernovae observed at Asiago in the seventies.
\thanks{Based on observations collected at ESO--La Silla, Chile.}}

\author{F. Patat\inst{1,4}, R. Barbon\inst{2}, E. Cappellaro\inst{3}
        and M. Turatto\inst{3}}
\institute{Dipartimento di Astronomia, Vicolo dell'Osservatorio 5, 
 35122 Padova, Italy \and
Osservatorio Astrofisico, 36012 Asiago (VI), Italy \and
Osservatorio Astronomico di Padova, Vicolo dell'Osservatorio 5, 
35122 Padova, Italy \and
ESO, K. Schwarzschild str. 2, D-85748 Garching bei M\"unchen, Germany}

\date{Received ................; accepted ................}

\maketitle
\markboth{F. Patat et al.}{Re-calibration of SN~Ia light curves}

\begin{abstract}
Following recent claims regarding possible errors in the photometry of SNe 
carried out at Asiago observatory during the 70's, which produced very 
blue $(B-V)$ color at maximum for some objects, we present
the result of new CCD photometry of the sequences around 16 type Ia 
supernovae. 

Except for a few cases, e.g. SN1970J and SN1972J for which a large
zero point error has been found, the new data show that the old Asiago
observations have been carried out properly and that their accuracy
was comparable to that expected for a photometry based on photographic
transfers.  This result is also substantiated by comparison with the
re-calibration of some Asiago sequences made photo-electrically by
Tsvetkov.

New light curves of the SNe  have been determined and 
B magnitudes and (B-V) colors at light peak derived.
With the new data the color distribution of the SNe studied here
becomes narrower and moves to the red by only 0.06 mag, showing no more 
very blue objects except for one, still uncertain, case.
As far as the use of SNIa as standard candles is concerned,
we show that the utilization 
of all SNe in the M$_B$ vs. (B-V)$^{max}_0$ plane
reduces the uncertainties due to the photometry. 

\end{abstract}


\section{Introduction}

The correction for reddening is a crucial step when using    SNe~Ia as
distance indicators.  To estimate the amount of reddening \ebv, suffered
by a given SN~Ia, two different approaches have been adopted
so far. 
One may infer the reddening through the use of a relation between the
intensity of the interstellar absorption lines and the total \ebv~
as suggested by Barbon et al. (1990).
Since this technique applies to SN spectra having sufficient
resolution and good signal to noise ratio, it has been used only for a
limited number of SNe (e.g. Patat et al. 1996).  Alternatively, the
extinction may be estimated from the observed SN color if one assumes
that all SN~Ia at a given phase, for instance at maximum light, have
the same intrinsic $(B-V)_0$ color. In this case the problem becomes
the determination of the intrinsic SN~Ia color at maximum.

 The most common approach is to adopt as reference the color
of the bluest objects in a given sample of SN~Ia, assuming that
they are those less affected by reddening. However this approach is
seriously endangered by possible photometric errors.  This may explain
the quite different intrinsic $(B-V)_0$ values which have been derived
in the past, e.g.: --0.20 (Pskovskii \cite{psk}), --0.15 (Barbon et
al. \cite{barb73b}), --0.27 (Cadonau et al. \cite{cadonau85}), --0.25
(van den Bergh \& Pazder \cite{vdbepaz}), --0.16 (Della Valle \&
Panagia \cite{mdv}), --0.15 (Branch \& Tammann \cite{bt92}), +0.03
(Branch \& Miller \cite{bm93}) and +0.09 (Sandage \& Tammann
\cite{st93}). As it can be seen, $(B-V)_0$ ranges about 0.35 mag with
recent estimates being, on the average, redder.

It is worth noting that since $A_B = R_B \times$ \ebv and $R_B\sim$4.0
(Savage \& Mathis \cite{savage}), even small errors in the adopted
$(B-V)_0$ reference color have strong impact on the derived
absolute magnitude $M_B$.

Since SN events are rare, one
is often forced to use data which, admittedly, are of poor photometric
quality in order to build a statistical significant sample.
If some
observations are affected by systematic errors, these may severely bias the
estimate of $(B-V)_0$.  In fact, several authors (Sandage \& Tammann
\cite{st93}, Branch \& Miller \cite{bm93}, Vaughan et al. \cite{vaugh},
Schaefer \cite{schaefer}) have noticed that most of the bluest SN~Ia
have been observed photographically at the Asiago Astrophysical
Observatory in the seventies and they raised the possibility that these
observations were affected by systematic errors. Since for most of
these "blue" SNe the only available photometry is that collected at
Asiago and bearing in mind the importance of SNIa in the determination
of the extragalactic distance scale, it appears of great interest to
check the accuracy of these old measurements.

The intensive program of photometric and spectroscopic monitoring of SNe
started at the Asiago Astrophysical Observatory in the early 60's. Before
the introduction, in the mid eighties, of CCD's the photometry of SNe was
obtained by photographic observations mainly using the 67/92~cm Schmidt
telescope and the 122 cm reflector. The SN magnitudes were estimated by
comparison with local sequences which, in turn, were calibrated by means of
photographic transfers from standard fields, usually photoelectrically
observed  Selected Areas.  To calibrate the local sequences, the
photographic plates were measured with a Becker iris photometer, whereas SN
magnitudes, especially for objects lying on the luminous background of
their parent galaxies, were often obtained by eye comparison with the local
sequences. 

Prompted by the aforementioned criticism, we decided to verify the old
Asiago photometry by re--calibrating the local comparison sequences
used to derive the SN magnitudes.  The comparison of the new
measurements with those reported in the original papers will provide
us a quantitative estimate of the reliability of the old SN data and,
in turn, will allow us to correct the light curves. We note that many
sequences were already re-calibrated by Tsvetkov at the Crimean Observatory
(for the individual
references see Table 1 and 2) who, however, used a photographic
technique similar to that adopted at Asiago for the majority of stars,
namely those fainter than B=14.5 mag.

To calibrate a significant sample of the local sequences, several nights of
good photometric quality are needed, which it is not presently the case at
Asiago due to the increasing light pollution and deteriorating sky
conditions. Therefore, among the SNe studied at Asiago, we selected those
accessible from ESO--La Silla ($\delta < +30^\circ$) which resulted in a
sample of 16 SNe covering about 25 years of observations
and requiring the re-calibration of about a hundred
stars of the local comparison sequences. 
These are listed in Tab.~\ref{tab:sample} whereas
Table 2 reports the SNe observed 
at Asiago not included in the present sample but with standard sequences 
 re--calibrated by Tsvetkov. Codes of the original papers are
reported in col.3 (A1-A9) along with those of the Tsvetkov's papers
(col.7, T1--T6). 
The passbands
of observations (Col.~4), the number of stars in the comparison
sequence (Col.~5), the number of photographic transfers
(Col.~6) are also reported.

\begin{table}
\tabcolsep 1.5mm
\caption{\label{tab:sample} SNe with re-calibrated local comparison sequences
.}
\begin{tabular}{llccccc}
      &            &       &         &     &     &     \\
\hline
SN     &  Galaxy   & Ref.*& Band    & N.S.& N.T.& Ref.* \\
\hline
1957B & NGC~4374     & A1  & $m_{\rm pg}$& 8 & 4  & T1\\
1960F & NGC~4496     & A1  & $m_{\rm pg}$& 5 & 1  & T2\\
1960R & NGC~4382     & A1  & $m_{\rm pg}$& 8 & 5  & T2\\
1965I & NGC~4753     & A2  & $B,V$     & 4 & 3  & T3\\
1968E & NGC~2713     & A2  & $B,V$     & 7 & 3  & T4\\
1970J & NGC~7619     & A3  & $B,V$     & 9 & 4  & T1\\
1971G & NGC~4165     & A3  & $B,V$     & 8 & 6  & T1\\
1972J** & NGC~7634     & A3  & $B,V$   & 9 & 4  & T1\\
1973B & Anon~1512+02 & A4  & $B$       & 3 & -- & --\\
1975N & NGC~7723     & A4  & $B,V$     & 4 & 6  & T1\\
1975O & NGC~2487     & A4  & $B,V$     & 6 & 7  & --\\
1976B & NGC~4402     & A4  & $B,V$     & 7 & 6  & --\\
1976J & NGC~977      & A4  & $B,V$     & 4 & 1  & --\\
1977F & M+5-26-14    & A4  & $B$       & 3 & 1  & --\\
1981B & NGC~4536     & A5  & $B,V$     & 8 & -- & T5\\
1983U & NGC~3227     & A6  & $B,V$     & 6 & -- & --\\
\hline
\multicolumn{7}{l}{NOTE: (*) coded as in the References}\\
\multicolumn{7}{l}{(**) same comparison sequence as for SN~1970J}\\

\end{tabular}
\end{table}

\begin{table}
\tabcolsep 2mm
\caption{\label{tab:sample2} SNe Ia observed at Asiago not included 
in our sample but re--calibrated by Tsvetkov.}
\begin{tabular}{llccccc}
      &            &       &         &     &     &     \\
\hline
SN     &  Galaxy   & Ref.*& Band   & N.S.& N.T.& Ref.* \\
\hline 
1969C & NGC~3811   & A7   &$B,V$       & 8 & 1  & T4\\
1971L & NGC~6384   & A3   &$B,V$       & 8 & 4  & T4\\
1972H & NGC~3147   & A8   &$B,V$       & 6 & 3  & T4\\
1972R & NGC~2841   & A8,A1&$B,V$       & 9 & 8  & T6\\
1973N & NGC~7495   & A9   &$B,V$       & 10& -- & T1\\
1974G & NGC~4414   & A9   &$B,V$       & 10& 5  & T1\\
1974J & NGC~7343   & A9   &$B,V$       & 7 & -- & T1\\
1975G & M+09-23-25 & A4   &$B,V$       & 4 & -- & T1\\
1975P & NGC~3583   & A4   &$B,V$       & 6 & -- & T1\\
\hline
\multicolumn{7}{l}{NOTE: (*) coded as in the References}\\
\end{tabular}
\end{table}

In the next Section we describe the observations and the data
reductions, in Sect.~3 we compare the revised magnitudes of the local
standard stars with those reported in the original papers, in Sect.~4 the
light and color curves of the SNe are re-calibrated and in Sect.~5 we
discuss the effects of the corrections on the SN color distribution.

\section{Observations and data reductions}

The observations have been conducted at ESO - La Silla during two
different runs, between March 7--12, and November 30 -- December 3,
1994.  B and V imaging of the sample stars have been obtained using
the 0.92~m ESO--Dutch telescope equipped with the TK coated 512
$\times$ 512 CCD \#29 in the first run and with the similar CCD \#33
in the second one.  The pixel scale was 0$^{\prime \prime}$.44 and the
field 3$^{\prime}$.3 $\times$ 3$^{\prime}$.3.  Typically the stars of
each local comparison sequence were spread about half a degree in the
sky hence it has been necessary to take several frames (up to 7, in
the case of SN~1971G) in order to measure all the stars of the
sequences. The brightness of the measured stars ranged from B$\sim 10$
down to B$\sim 19$ mag and accordingly the exposure times ranged from 3 secs up
to 20 min.

In addition to the program stars, a number of photometric standard
fields (Landolt \cite{landolt}) has been frequently monitored during
the two runs.  Most of the nights were photometric and the few
observations obtained in non--photometric conditions have been
rejected. In a number of cases, the same program star has been observed
several times in the same or different nights and the observed magnitudes 
have been averaged.

The frames have been flat--fielded and bias corrected by means of
standard routines in the ESO--MIDAS environment.  In all cases the
stars were located relatively far from the SN parent galaxy, and it was
therefore safe to measure the star flux by means of plain aperture
photometry with the background estimated in a circular annulus
centered on the star. The diaphragm size was chosen according to the
seeing, which ranged from $1^{\prime \prime}$ to $2^{\prime \prime}$
(FWHM).

The instrumental magnitudes have been transformed into B and V
magnitudes using the fitting coefficients derived from the
observations of the standard fields, after
including airmass correction ($K_B$=0.27, $K_V$=0.12 mag
airmass$^{-1}$).

The transformation equations turned out to be:

\begin{eqnarray}
&B = C_B + b + 0.10 (\pm 0.01) \times (B-V)\\
&V = C_V + v + 0.03 (\pm 0.01) \times (B-V)
\end{eqnarray}

where $b$, $v$ are the instrumental magnitudes corrected for
atmospheric extinction, $C_B$ and $C_V$ are the zero points showing
small variation from night to night and the color terms were averaged
on all the nights (in parenthesis is their r.m.s. scatter). 

To estimate the internal errors, we performed artificial star
experiments by introducing, in the original frames, stars of known
magnitudes: the standard deviation of the magnitude of the recovered
artificial stars was found 0.01 mag at B=12.0 and 0.05 mag at B=18.0.

It is worthwhile to note that recently Schaefer (\cite{schaefer96})
has re-calibrated, also using a CCD detector, the sequence of
SN~1960F. His estimates of the magnitudes of the sequence stars are
always within 0.05 mag from those presented in this work.

\section{\label{par:ric} Re-calibration of the local sequences}

The re-calibrated CCD magnitudes for the program stars included in the
sample are reported in Tab.~3. $B$ magnitudes are in column 3
and $B-V$ colors in column 4. For comparison, in column 5
and 6 the original estimates are also reported. Notice that these latter 
ones for
 SNe~1957B, 1960F and 1960R were
in the $m_{\rm pg}$ system (Bertola \cite{bertola64}).

\begin{table*}
\tabcolsep 2.0mm
\caption{\label{tab:data} Re--calibrated $B^*$ magnitudes and $(B-V)^*$ colors
 for the stars of the local sequences. For comparison , also the
Asiago original data are given. Identification charts are found in the
original papers (Table~1).}
\begin{tabular}{llcccc|llcccc}
\hline
   & &
\multicolumn{2}{c}{this work} &
\multicolumn{2}{c|}{Asiago photographic} &
   & & 
\multicolumn{2}{c}{this work} &
\multicolumn{2}{c}{Asiago photographic} \\
SN & star& $B^*_{\rm CCD}$ &$(B-V)^*_{\rm CCD}$& $B^*_{\rm As}$ &
$(B-V)^*_{\rm As}$ & SN & star& $B^*_{\rm CCD}$ &$(B-V)^*_{\rm CCD}$&
$B^*_{\rm As}$ & $(B-V)^*_{\rm As}$\\
\hline
57B$^\#$& a & 12.41 & 0.49 & 12.40 &   &73B&a& 17.56 & 0.93 & 17.40 & 0.80\\
    & b & 13.86 & 0.90 & 13.83 &     &   &b&  \#\#  &      & 18.20 &    \\     
    & c & 14.20 & 0.52 & 14.20 &     &   &c& 16.84 & 0.93 &       &    \\ 
    & e & 15.66 & 0.71 & 15.45 &     &   &d& 18.75 & 1.39 &       &    \\
\cline{7-12}
    & f & 15.95 & 0.85 & 15.75 &     &75N&c& 14.68 & 0.56 & 14.40 &0.60\\
    & g & 16.39 & 0.57 & 16.06 &     &   &e& 15.33 & 0.71 & 15.10 & 0.70\\
    & h & 16.55 & 1.08 & 16.24 &     &   &g& 16.37 & 0.73 & 15.90 & 0.65\\
    & i & 16.93 & 0.81 & 16.68 &     &   &h& 17.14 & 0.83 & 16.75 & 0.45\\
\cline{1-6} \cline{7-12}
60F$^\#$& a &  9.72 & 0.53 & 9.76 &    &75O&b& 15.64 & 0.83 & 15.50 & 0.70\\ 
    & b & 12.26 & 0.88 & 12.45 &     &   &c& 16.09 & 0.65 & 16.10 & 0.40\\
    & c & 12.68 & 0.69 & 12.73 &     &   &d& 16.85 & 0.74 & 16.95 & 0.70\\
    & d & 14.42 & 0.77 & 14.53 &     &   &e& 16.99 & 0.50 & 17.25 & 0.45\\
    & e & 15.29 & 0.76 & 15.50 &     &   &f$^{\#\#\#}$& 14.97 & 0.65 & 17.95 & 1.30\\
\cline{1-6}
60R$^\#$& a & 12.37 & 0.62 & 12.45 &   &   &g& 17.90 & 0.69 & 18.15 & 0.85\\ 
\cline{7-12}
    & b & 13.58 & 0.72 & 13.40 &     &76B&a& 14.54 & 0.77 & 14.20 & 0.60\\
    & c & 14.30 & 0.67 & 14.10 &     &   &c& 15.05 & 0.59 & 14.80 & 0.40\\ 
    & d & 14.76 & 0.74 & 14.50 &     &   &d& 15.20 & 0.78 & 14.95 & 0.70\\ 
    & e & 14.87 & 0.78 & 14.68 &     &   &e& 15.88 & 1.08 & 15.55 & 1.00\\
    & f & 15.24 & 0.70 & 14.92 &     &   &g& 16.44 & 0.72 & 16.20 & 0.50\\ 
    & g & 16.40 & 0.07 & 15.94 &     &   &h& 16.87 & 0.69 & 16.80 & 0.60\\ 
    & h & 17.07 & 0.78 & 16.25 &     &   &i& 17.12 & 0.62 & 17.40 & 0.45\\
\cline{1-6} \cline{7-12}
65I & a & 12.47 & 1.23 & 12.20 & 1.50&76J&a& 14.36 & 0.62 & 15.00 & 0.80\\
    & b & 13.41 & 1.10 & 13.40 & 1.30&   &b& 15.00 & 0.66 & 15.90 & 1.00\\
    & c & 13.76 & 0.72 & 13.40 & 1.30&   &c& 15.75 & 0.83 & 16.60 & 0.70\\
    & d & 14.34 & 0.64 & 14.25 & 0.58&   &d& 18.16 & 0.68 & 17.80 &\\
\cline{1-6} \cline{7-12}
68E & a & 14.97 & 0.53 & 14.90 & 0.34&77F&a& 16.39 & 0.64 & 16.10 &\\ 
    & b & 15.38 & 0.71 & 15.47 & 0.56&   &b& 17.99 & 0.84 & 18.00 & \\
    & c & 15.75 & 0.95 & 15.75 & 0.75&   &b& 18.61 & 1.51 & 18.50 &\\
\cline{7-12}
    & d & 15.92 & 0.54 & 15.93 & 0.23&81B&a& 10.22 & 1.00 & 10.06 & 0.95\\
    & e & 16.29 & 0.72 & 16.30 & 0.42&   &b& 12.16 & 0.49 & 12.37 & 0.51\\
    & f & 16.86 & 0.66 & 16.93 & 0.48&   &c& 12.61 & 0.94 & 12.75 & 0.90\\
    & g & 16.99 & 0.79 & 17.12 & 0.48&   &d& 13.77 & 0.80 & 13.85 & 0.82\\
\cline{1-6}
70J--72J&a&14.51& 0.71 & 13.85 & 0.25&   &e& 14.25 & 0.96 & 14.33 & 1.04\\
    & b & 14.99 & 0.68 & 14.35 & 0.25&   &f& 14.36 & 0.23 & 14.42 & 0.32\\
    & c & 15.81 & 0.60 & 15.15 & 0.20&   &g& 15.25 & 0.84 & 15.29 & 0.86\\
    & d & 16.27 & 0.59 & 15.50 & 0.15&   &i& 16.05 & 0.83 & 15.98 & 0.91\\
\cline{7-12}
    & e & 16.71 & 0.52 & 15.90& --0.05&83U&1& 13.46 & 0.43 & 13.35 & 0.40\\
    & f & 16.99 & 0.64 & 16.25 & 0.20&   &2& 14.14 & 0.67 & 14.25 & 0.75 \\
    & g & 17.41 & 0.71 & 16.85 & 0.60&   &3& 14.66 & 1.01 & 14.85 & 1.20\\
    & h & 17.91 & 0.50 & 17.30 & 0.00&   &4& 15.41 & 0.62 & 15.55 & 0.75\\
    & i & 18.62 & 1.57 & 17.80 & 1.20&   &5& 15.87 & 0.62 & 15.95 & 0.60\\
    & l & 18.74 & 1.54 & 18.25 & 1.50&   &6& 16.80 & 0.62 & 16.90 & 0.85\\
\cline{1-6}
71G & a & 13.87 & 0.95 & 13.60 & 0.85&   & &       &      &       & \\
    & b & 14.56 & 0.83 & 14.30 & 0.75&   & &       &      &       &\\
    & c & 14.94 & 0.65 & 14.65 & 0.40&   & &       &      &       &\\
    & d & 15.16 & 0.93 & 15.05 & 1.20&   & &       &      &       &\\
    & e & 16.20 & 0.94 & 16.15 & 1.05&   & &       &      &       &\\
    & f & 16.62 & 0.41 & 16.60 & 0.60&   & &       &      &       &\\
    & g & 17.33 & 0.53 & 17.15 & 0.70&   & &       &      &       &\\
    & h & 17.88 & 0.51 & 17.65 & 0.90&   & &       &      &       &\\
\hline
\end{tabular}

\# Original photometry in the $m_{\rm pg}$ system. \\
\#\# The CCD image shows that this is a galaxy, not a star.\\
\#\#\# Because of the large discrepancy ($B^*_{\rm CCD}-B^*_{\rm As}=2.98$ mag)
most likely the star was misidentified on the published map (Ciatti \&
Rosino 1978). For this reason it has been excluded from the
discussion.
\end{table*}

In principle, two types of errors can affect photometric data: a)
systematic errors possibly related to the reproduction of the
photometric system; b) random errors dependent on the detector type,
method of measurement etc. The calibration of the zero point of the
photometric system for each sequence is also affected by errors due to the
photographic transfer.  As a first step we want to verify that the
latter ones are random errors. 

To this purpose we calculated, for each sequence reported in Table~3,
the average zero--point offset of the calibration of the $B$
magnitudes and of the $(B-V)$ colors, $\Delta B^* = < B^*_{\rm As} -
B^*_{\rm CCD} >$ and $\Delta (B-V)^* = < (B-V)^*_{\rm As} -
(B-V)^*_{\rm CCD} >$ and the relative dispersions. The results are
reported in Tab.~4 in which at the bottom line are given the averages
of these quantities on all sequences. The average magnitude and color
offsets, $<\Delta B^*> = -0.09\pm 0.07$ and $<\Delta (B-V)^*> =
-0.05\pm 0.06$ respectively, show no evidence of systematic effects in
the transfer of the magnitudes of the local sequences.  Moreover, the
average dispersion $<\sigma_B^*>=0.17$ gives a good estimate of the
internal errors of the Asiago original photometry, assuming that the
errors of the CCD photometry are negligible.

As we mentioned in the introduction, in a series of papers Tsvetkov
(1982, 1983, 1985a, 1985b, 1986, 1996) 
re-calibrated many of the original Asiago comparison sequences.
His photometry was obtained using a
EMI photomultiplier coupled to the 60 cm telescope (Crimean
Observatory) for stars brighter than $V$=14.5 and using a  
15$^{\prime \prime}$ or 26$^{\prime \prime}$
diaphragm, according to the seeing. Fainter stars were
measured on photographic plates taken with the 50 cm Maksutov
telescope using a Racine wedge.  Many of these sequences, those
accessible from La Silla, have been re-calibrated also by us (see Tab.~1).
Since
in the following we will refer also to these data, it is of interest to
verify how they compare with our more accurate CCD estimates.  It turns out
that the average magnitudes and color offsets between Tsvetkov's data and
those of this work, $<\Delta B^*> = < B^*_{\rm Tsv} -
B^*_{\rm CCD} > =
 -0.02\pm 0.04$ and $<\Delta (B-V)^*> = < (B-V)^*_{\rm Tsv} -
(B-V)^*_{\rm CCD} > =
-0.04\pm 0.02$, are negligible. In analogy to what was done in Tab.~4, we
derived also an estimate of the internal error of the Tsvetkov
photometry,  $<\sigma_B^*>=0.10$, therefore about half that one of the
Asiago photometry.  In conclusion, even if the Tsvetkov photometry has
not the same accuracy as the new CCD estimates, we will use it to
check the Asiago photometry of the sequences we could not measure in
La Silla.
For these sequences, the Asiago magnitude and color offsets, relative
to the re-calibration by Tsvetkov, are reported in Tab.~5.

\begin{table}
\caption{\label{tab:delta} Zero point offsets, $\Delta B^*$ and
$\Delta (B-V)^*$, and dispersions of the original photometry with
respect to the CCD data for each of the re-calibrated sequences.}
\tabcolsep 1.5mm
\begin{tabular}{lcccc}
         &           &     &          &           \\
\hline
SN       &$\Delta B^*$&$\sigma^*_B$& $\Delta(B-V)^*$&$\sigma^*_{B-V}$ \\
\hline
1957B$^\#$  & $-0.17$  &  0.14  & --    &        \\
1960F$^\#$  & $+0.12$  &  0.08  & --    &        \\
1960R$^\#$  & $-0.29$  &  0.26  & --    &        \\
1965I       & $-0.17$  &  0.16  &$+0.25$&  0.26  \\
1968E       & $+0.03$  &  0.07  &$-0.22$&  0.06  \\
1970J--72J  &$-0.67$   &  0.11  &$-0.38$&  0.17  \\
1971G       & $-0.17$  &  0.10  &$+0.08$&  0.21  \\
1973B       & $-0.16$  &  --    &$-0.13$&   --   \\
1975N       & $-0.33$  &  0.11  &$-0.11$&  0.19  \\
1975O       & $+0.10$  &  0.17  &$-0.12$&  0.10  \\
1976B       & $-0.16$  &  0.22  &$-0.14$&  0.06  \\
1976J       & $+0.51$  &  0.59  &$+0.13$&  0.24  \\
1977F       & $-0.12$  &  0.15  & --    &      \\
1981B       & $+0.05$  &  0.15  &$+0.03$&  0.05  \\
1983U       & $+0.09$  &  0.10  &$+0.10$&  0.11  \\
\\
	& $< B^* >$& $< \sigma^*_B >$ & $< \Delta(B-V)^* >$& $<
\sigma^*_{B-V} >$ \\
            &$-0.09$   &$0.17$  &$-0.05$&  0.15 \\ 
            &$\pm 0.07$&      &$\pm 0.06$ &    \\
\hline 
\end{tabular}

\# Original photometry in the $m_{pg}$ system.
\end{table}

\begin{table}
\caption{\label{tab:delta2}Zero point offsets and dispersions of the original
photometry with respect to Tsvetkov's re--calibration of the 
sequences of Table~2.}
\tabcolsep 1.5mm
\begin{tabular}{lcccc}
        &           &     &          &           \\
\hline
SN  &   $\Delta B^*$ & $\sigma^*_B$ & $\Delta (B-V)^*$ & $\sigma^*_{B-V}$ \\
\hline  
1969C~~~~~ & $+0.12$  & 0.19 & $+0.03$ &  0.28  \\
1971L & $+0.28$  & 0.10 & $+0.17$ &  0.32  \\
1972H & $ 0.00$  & 0.14 & $-0.20$ &  0.06  \\
1972R & $+0.13$  & 0.12 & $-0.19$ &  0.18  \\
1973N & $+0.25$  & 0.13 & $+0.37$ &  0.22  \\
1974G & $-0.08$  & 0.09 & $-0.15$ &  0.16 \\
1974J & $+0.07$  & 0.12 & $+0.19$ &  0.14  \\
1975G & $+0.21$  & 0.13 & $-0.12$ &  0.13 \\
1975P & $-0.12$  & 0.12 & $+0.02$ &  0.11  \\
\\	
	& $< B^* >$& $< \sigma^*_B >$ & $< \Delta(B-V)^* >$& $<
\sigma^*_{B-V} >$ \\
      &$+0.10$&$0.13$  &$+0.01$&  0.18 \\ 
          &$\pm 0.05$&      &$\pm 0.07$ &    \\
\hline
\end{tabular}
\end{table}

\begin{figure}
\psfig{figure=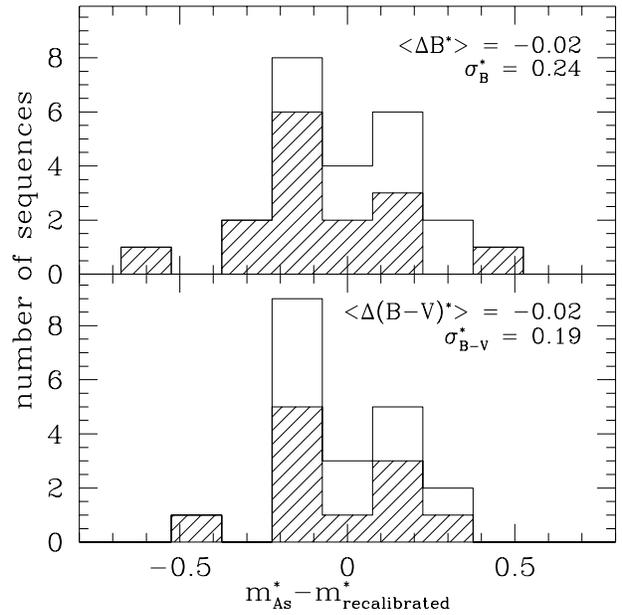,width=8.5cm}
\caption{ Upper panel: distribution of the zero point offsets $\Delta
B^*$ of
the Asiago photometry compared with CCD re-calibrated estimates 
(shaded area)
and with Tsvetkov's estimates (unshaded area). 
Lower panel: the same for the $\Delta (B-V)$ colors.}\label{fig:hist}
\end{figure}

For the combined sample of Tables 4 and 5, the distribution of the
zero point offsets is displayed in Fig.~\ref{fig:hist}. It is
confirmed what stated before: the distributions of the magnitude and
color offsets are quite broad, indicating that the random errors in
the calibration of the zero points are relatively large, the average
error being $\sim 0.2$ mag. On the contrary, there are no evidences of
systematic errors in the calibration both for the $B$ magnitudes and
the $(B-V)$ colors.

Clearly, a random error in the zero point calibration of a given
sequence translates in a systematic error for the SN light curve.
This turns out to be especially severe for the SNe 1970J and 1972J,
calibrated on the same sequence, and for SN~1976J.  The latter case is not
surprising since in the original paper (Ciatti \& Rosino 1978) it was
stated that the sequence of SN~1976J was calibrated by only one
photographic transfer and therefore the results ``must be considered very
preliminary''. 

More cumbersome is the case of SNe 1970J and 1972J for which we could not
find an obvious explanation for the very large calibration error. A
check of the Asiago Observatory archive has shown that the calibration
of the comparison sequence was obtained through 4 photographic
transfers from SA~68, which was observed in the same nights and at
similar zenith distance as the SN field.  Another
possibility we investigated was that of errors in the reference
magnitudes of the stars of SA~68 (Stebbins et al. \cite{stebbins}). To
our knowledge, the only recent photometry of the SA~68 stars was
published by Doroshenko (\cite{doroshenko}), who found small
differences with respect to the values of Stebbins et al. 
(note that the different star identifications may be
confusing). The comparison sequences of the other SNe included in
the sample which have been calibrated with transfers from SA~68,  do
not show the severe errors of the sequence of SN~1970J. Therefore the
origin of the errors affecting SNe 1970J and 1972J remains unknown.

\begin{figure}
\psfig{figure=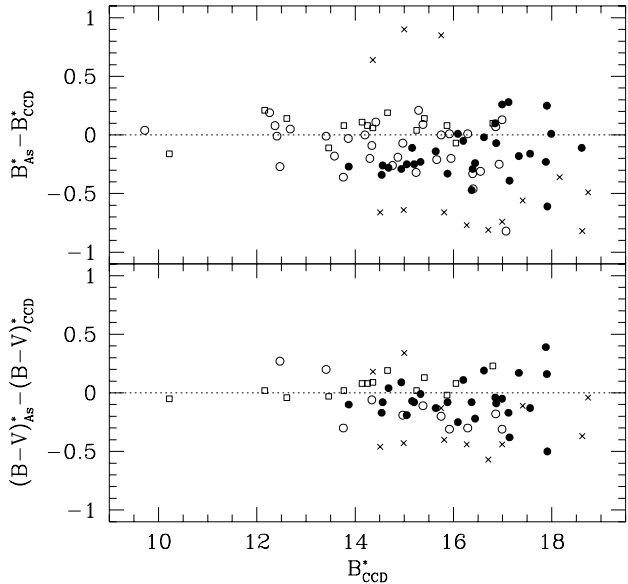,width=9cm}
\caption{\label{fig:as}  Deviations of the Asiago $B$ magnitudes
(upper panel) 
and $(B-V)$ colors (lower panel) from the new CCD photometry for
the stars of the local sequences.  Abscissa are $B^*_{\rm CCD}$
magnitudes. Different symbols have been used to discriminate  the Asiago
measurements of three different decades, namely: 1957-1969 open
circles, 1970-1979 filled circles, 1980-1983 open square. The two most
deviating sequences, i.e.  those of SNe 1970J-1972J and SN~1976J, are
indicated by crosses.}
\end{figure}

In principle, systematic errors may correlate with the brightness of
the stars or with their colors. 
To investigate the first possibility, in Fig.~\ref{fig:as} we plot the
differences $B^*_{\rm As} - B^*_{\rm CCD}$ and $(B-V)^*_{\rm
As}-(B-V)^*_{\rm CCD}$ as a function of $B^*_{\rm CCD}$ for all of the
stars of Tab.~3. As it can be seen, the dispersion of the
Asiago photometry increases as the sequence stars become fainter and it
might also be present a slight dependence on magnitude, mainly induced by
the sequences measured in the seventies which were, in the average, fainter
and therefore more affected by random errors.

\begin{figure*}
\psfig{figure=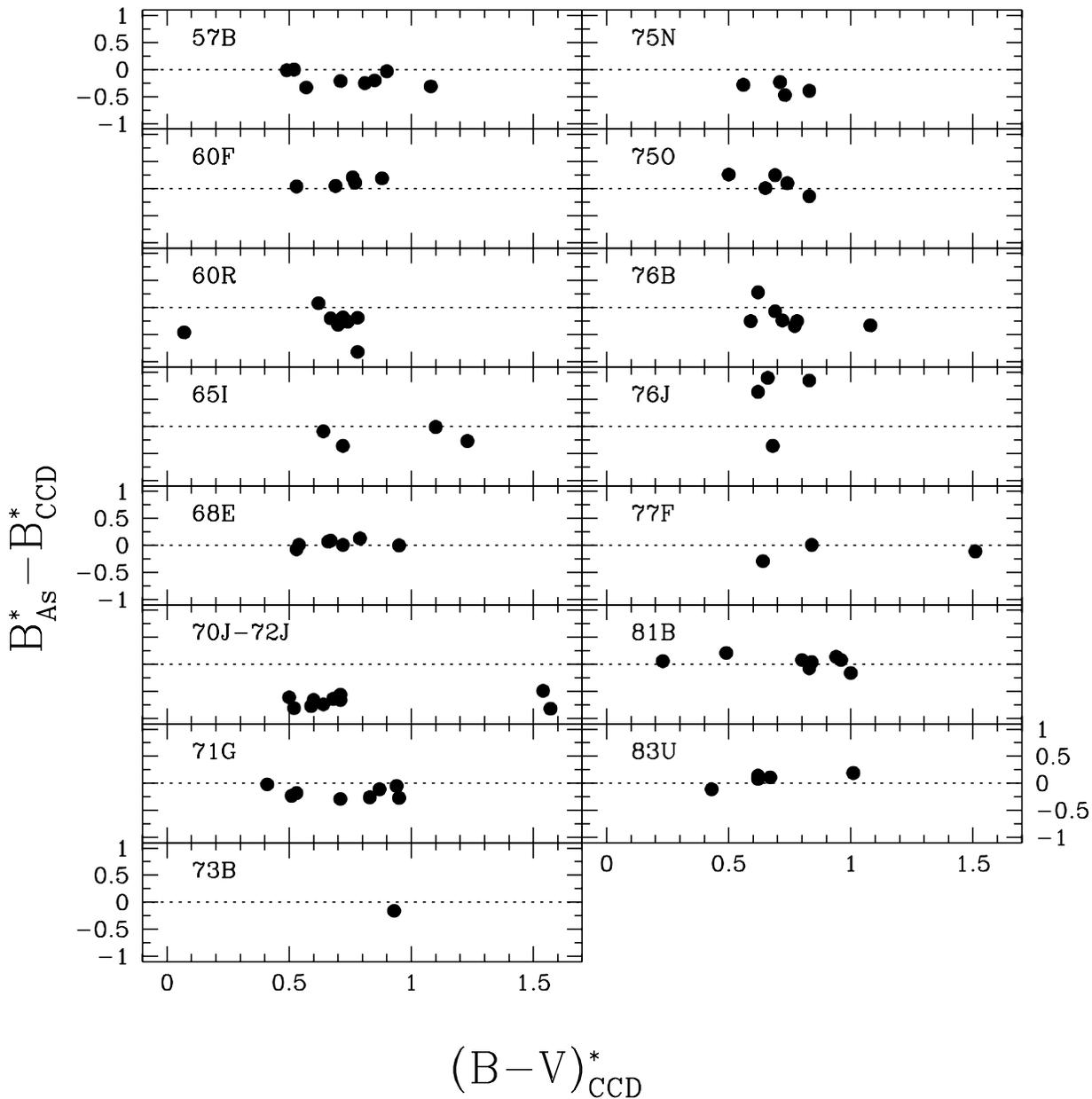,width=18cm}
\caption{ Deviations of the Asiago 
photometry from the new CCD photometry as a function of the
re-calibrated $(B-V)^*_{\rm CCD}$ color. The scale is the same for all
panels.}\label{fig:colors}
\end{figure*}

In the literature (e.g. Sandage \& Tammann \cite{st93}) 
it has been raised the doubt that some of
the Asiago photometry was obtained without the proper plate-filter
combination. A check of the Asiago plate archive confirmed that B
and V photometry were always obtained using Kodak 103a--O (or IIa--O)
plates and the
GG~13 filter and 103a--D+GG~14, respectively.  The use of a wrong
plate+filter combination would have produced a color dependence
of the errors of the photometry. To check this
point in Fig.~\ref{fig:colors} we plotted, for each sequence, the
deviations of the B original photometry from the new one, as a function of
the re--calibrated $(B-V)$ color.   
It results that, even when the deviations are
large, as in the sequences of SNe 1970J-1972J and 1976J, there are no
evidences of any dependence on the $(B-V)$ colors. 

\section{Re--calibration of the light curves}

We have seen that, for each sequence, there is a significant dispersion
around the zero point offset (col.3 of Tab.~4), that is the magnitude
corrections were different even for stars of the same sequence. Since the SN
magnitudes, at each phase, have been obtained by interpolation between the
couple  of comparison stars close in brightness to
the SN,  the correction of the
magnitudes and colors  of the SN  is not simply a zero point correction. 

In principle, to re--draw the SN light curves it would be better
to derive new estimates of the SN magnitudes against the re--calibrated local
sequence stars, possibly using modern technique, e.g. digitizing the
plates with a PDS machine and measuring the SN intensities with
suitable software.  However our experience show that: a) the internal
errors originally made in the comparison were small ($\sim 0.1$ mag)
compared to the errors of the calibration of the sequence; b) there
is an intrinsic limit for the photometric accuracy which can be
achieved using photographic plates, in particular those obtained with
our Schmidt telescope ($\sim 0.05$ mag); c) the procedure is time
consuming implying the digitizing of hundreds of plates.

For these reasons we decided to correct the SN photometry adopting the
original estimates of the SN magnitudes relative to the comparison
stars, that is, if $m_{SN}$ is the original SN measure, $m_A$, $m_B$
the original magnitudes of the comparison stars ($m_A \leq m_{SN}
\leq m_B$) and $m'_A$, $m'_B$  the re-calibrated magnitudes
for the same stars, the corrected SN magnitude $m'_{SN}$ is obtained
through the relation:

\begin{equation}
 m'_{SN} = m'_A + (m_{SN}-m_A) \times \frac{\Delta m'}{\Delta m}
\end{equation}
 
where $\Delta m = (m_B-m_A)$ and $\Delta m' = (m'_B-m'_A)$. This
relation is correct only if the ranking in the comparison sequence
remain the same, that is $m'_a < m'_b < m'_c .....$. Moreover, it is not
safe to apply the formula if $\Delta m'$ is very different from
$\Delta m$. In these cases, to correct  the SN magnitude, we
just referred to one comparison star, namely that closest in
brightness to the SN. This is equivalent to putting $\Delta m$ = $\Delta
m'$ in Eq.~3. The same approach is followed in the few cases in which one
of the two stars of the comparison sequence have not been
re--observed. In these and in cases in which
extrapolation is needed, the result are considered uncertain.

We remind that the original photometry of SNe 1957B, 1960F and 1960R
was in the $m_{pg}$ system, whereas the stars of the sequences were
re-calibrated in the B system. 
Second order corrections due to the color terms resulted smaller than
the accuracy of the measurements, therefore they were neglected.
Following these prescriptions, the photometric observations of all the
SNe listed in Tab.~\ref{tab:sample} have been
re--calibrated. 

\begin{figure}
\psfig{figure=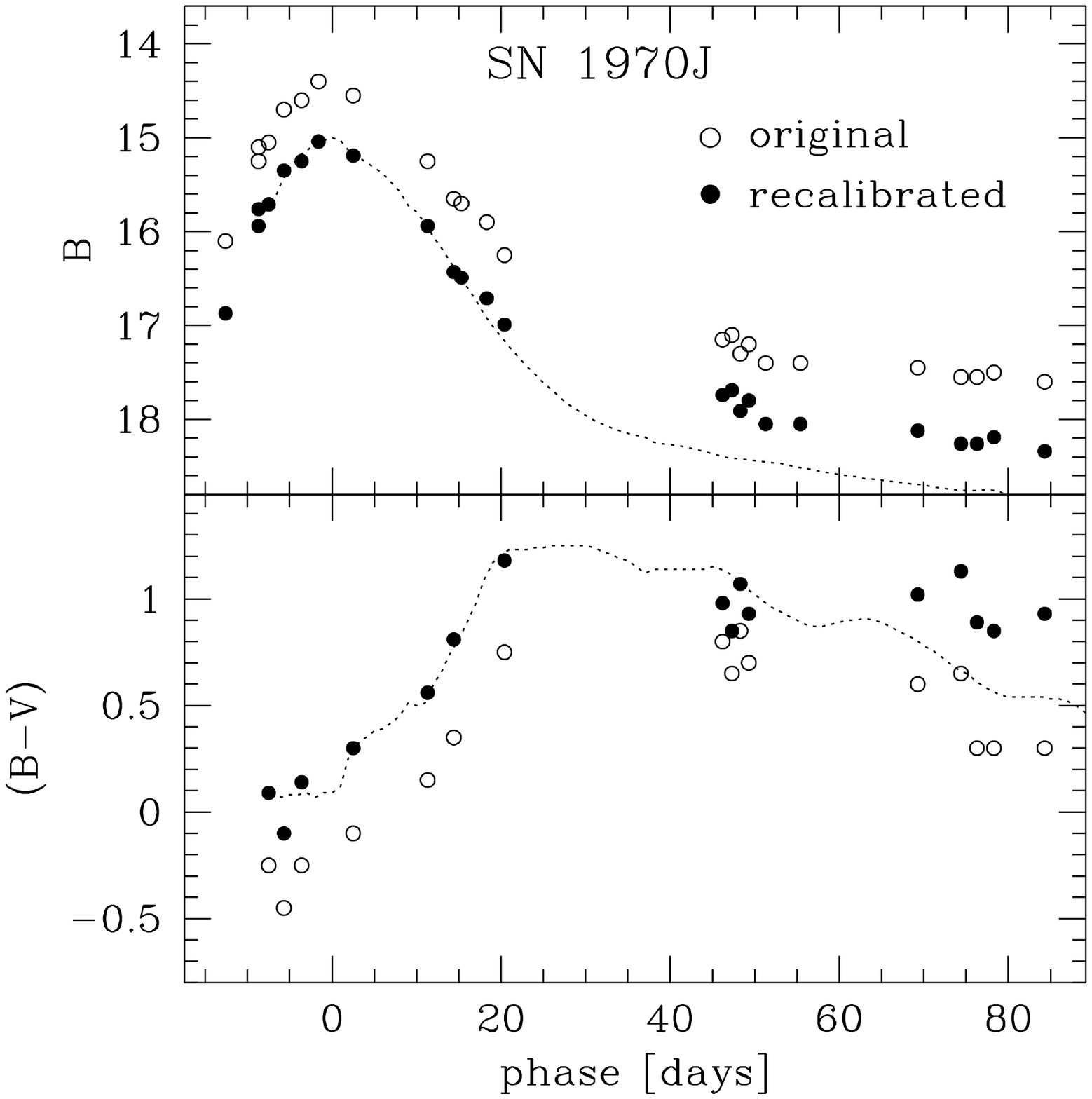,width=9cm}
\caption{ Example of re-calibration: light and color curves of 
SN~1970J.  Overimposed (dotted line) is a best fit to the light and
color curves of SN~1992A (Suntzeff 1995).}\label{fig:70j}
\end{figure}

Having re-calibrated the old SN light curves, we can now derive new estimates
of $B$ and $(B-V)$ at maximum light.  Generally, these have been
obtained by a best fit to the $B$ and $B-V$ light curves of the typical
SN~Ia SN~1992A (Suntzeff \cite{suntzeff}, Kirshner et al. \cite{kirsh}).
Whereas in most cases the corrections to be applied are fully consistent
with the estimates of the errors reported in the original papers ($\sim
0.2$ mag), there are cases in which the deviations are severe. 

The first case is that of SN~1970J which, on the basis of the original
photometry was
estimated $\sim 0.7$ mag and $\sim 0.3$ mag brighter in the B and V,
respectively, and this resulted in a $(B-V)_{\rm max} \sim 0.4$ mag
too blue.
Fig.~\ref{fig:70j} shows the new light and color curve of this SN as
an example of the new re-calibrations.  
There is another source of photometry for this
SN (Dubjago \& Tokhtsev \cite{dubjago}), which is in agreement with the
original Asiago data. This is not surprising since they used the 
same local sequence and it demonstrates
that at least the Asiago estimates of the SN with respect to the sequence
were correct.  It is worth to note
that at late phases, the re-calibrated B light curve appears $\sim$0.6 mag
brighter than the template. This effect, which is found in many  other
photographic light curves, is probably
due to errors related to the poor contrast
of the SN against the parent galaxy background (cfr. Boisseau \&
Wheeler \cite{bois}). Unfortunately, the
comparison sequence of SN~1970J was used also to estimate the magnitudes of
SN~1972J which, in turn, resulted affected by the same errors.

Also SN~1975O had a $(B-V)_{\rm max}$ color $\sim$0.3 mag too blue
in the original photometry compared with the re-calibrated value. In
this case, the difference is not due to a systematic calibration error
of the sequence, but to random
errors in the V magnitude of stars $b$ and $c$ which conspired to give
different extrapolated maxima.

We can also re-calibrate the SNe which sequences have been re-measured
by Tsvetkov, using the same procedure described above.  SN~1972H turns
out to be $\sim$0.4 mag redder in proximity of maximum light. The fit
to the templates was quite uncertain, due to the fact that the first
observation has been obtained two weeks after maximum.  The color
drift is due to the revised V magnitudes of stars $a$ and $b$.  Small
variations are seen in the early observations of SN~1973N, but around
two months after maximum, when the SN was close to the plate limit
($B\sim18.5$), the B magnitudes were too faint by $\sim0.2$ mag and the Vs were
too bright by the same amount.  The new fittings give a $(B-V)_{\rm
max}$ bluer by 0.4 mag with respect to previous estimates.

Puzzling is also the case of SN~1974J.   The original
photometry indicated a fairly blue color at maximum ($(B-V)_{\rm max} =
-0.28$) raising, as stressed by Sandage \& Tammann (\cite{st93}) and
Vaughan et al. (\cite{vaugh}),  doubts on the photometry. The
re-calibrated light curves are even bluer, with a  revised color estimate
$(B-V)_{\rm max} = -0.4$ mag. Both the original and the re-calibrated
light curves show a very flat peak, especially in the B band, which clearly
deviates from the behavior of the template and  causes  the fit to be 
very uncertain.  However, we remind that the errors in the Tsvetkov
re-calibration, although smaller that those of the original Asiago
photometry are not negligible (see Tab.~5).  Only CCD re-calibration of the
sequence and a new reduction of the original plates may clarify if something
is wrong, if any, with this SN. 

The revised estimates of $B_{\rm max}$ and $(B-V)_{\rm max}$, are reported
in Tab.~\ref{tab:resu}, along with the estimated errors on $(B-V)_{\rm max}$ as
deduced by the fitting uncertainties. From this Table, SNe 1968E, 1973B and
1983U have been excluded owing to the poor accuracy of the fits to the
template. Finally, in the last column we report the  color shift between
the old  $(B-V)_{\rm max}$ from Leibundgut et al. (1991),
adding the galactic reddening,and the revised values. 

\begin{table}
\caption{\label{tab:resu} Revised $B_{\rm max}$ and $(B-V)_{\rm max}$
(not corrected for galactic absorption) for the SNe of Table 4 and 5,
after sequence re-calibration. 
In column 3 are  estimates of the fitting
uncertainties on the $(B-V)_{\rm max}$ color.
In the last column are
the differences between the original colors and the revised
ones.}
\begin{tabular}{lccc}
\hline
SN    & $B_{\rm max}$ & $(B-V)_{\rm max}$ & $\Delta(B-V)_{\rm max}$ \\
\hline
\multicolumn{4}{c}{this work}\\
1957B & 12.20   &  --        &  --  \\ 
1960F & 11.34   &  --        &  --  \\
1960R & 11.60   &  --        &  --  \\
1965I & 12.41   &$-0.19\pm 0.30$  & $-0.04$ \\
1970J & 15.00   &$+0.12\pm0.05$& $-0.45$ \\
1971G & 13.90   &$-0.10\pm0.30$& $+0.10$ \\
1972J & 14.76   &$+0.11\pm0.05$& $-0.38$ \\
1975N & 14.00   &$+0.30\pm0.30$& $-0.13$ \\
1975O & 15.30   &$+0.14\pm0.30$& $-0.32$ \\
1976B & 15.10   &$+1.24\pm0.30$& $-0.07$ \\
1976J & 14.28   & $0.00\pm0.20$& $-0.13$ \\
1977F & 15.80   &   --       &   --    \\
1981B & 11.74   &$+0.06\pm0.05$& $-0.09$ \\
\\
\\
\multicolumn{4}{c}{Tsvetkov's re-calibration}\\
1969C & 13.79   &$+0.05\pm0.20$& $+0.07$\\
1971L & 13.00   &$+0.25\pm0.30$& $+0.23$\\
1972H & 14.40   &$+0.15\pm0.30$& $-0.36$\\
1972R & 12.85   &$+0.05\pm0.30$& $-0.17$\\
1973N & 14.91   &$-0.04\pm0.30$& $+0.43$\\
1974G & 12.28   &$+0.30\pm0.20$& $-0.13$\\
1974J & 15.60   &$-0.40\pm0.30$& $+0.18$\\
1975G & 14.44   &$-0.15\pm0.30$& $+0.15$\\
1975P & 14.44   &$+0.26\pm0.40$& $-0.09$\\
\hline
\end{tabular}
\end{table}

\section{Absolute magnitude and color distributions}

From the data of Tab.~\ref{tab:resu} we can derive revised estimates
of the absolute magnitudes and corrected colors to be compared with
previous estimates for the same SNe and with other SN~Ia samples.  To
this purpose we used distance moduli $\mu$ from Vaughan et
al. (\cite{vaugh}) or, if not available, from Tully (\cite{tully}) or
LEDA\footnote{The Lyon--Meudon Extragalactic Database (LEDA) is
supplied by the LEDA team at the CRAL--Observatoire de Lyon (France)}
normalized to $H_0=85$ km s$^{-1}$ Mpc$^{-1}$ for consistence with
Vaughan et al.  (\cite{vaugh}).  The galactic reddening $A_B$ was
taken from Burstein \& Heiles (\cite{burst}).  Instead, no correction
has been applied for the possible reddening within the parent
galaxies. SNe 1957B, 1960F and 1960R were excluded because the
original plates were taken in the $m_{\rm pg}$ system and because no
color is available.

The resulting $M_B$ absolute magnitudes and the $(B-V)_0^{\rm max}$
colors are reported in Tab.~\ref{tab:max}, and compared to the
previous estimates. These latters have been derived from the peak 
magnitudes reported by Leibundgut et al. (1991) and using the distance 
moduli listed in Table~\ref{tab:max}.

\begin{table*}
\caption{ Absolute magnitudes and colors for the SNe
in the sample corresponding to $H_0=85$ km s$^{-1}$ Mpc$^{-1}$.
In the last two columns are the previous estimates derived from
Leibundgut et al. (1991) data.}\label{tab:max}
\begin{center}
\begin{tabular}{lcccccc}
\hline
\multicolumn{3}{c}{ } &
\multicolumn{2}{c}{Re-calibrated} &
\multicolumn{2}{c}{Previous estimates} \\
SN     &$A_B$&$\mu$&$M_B$ &$(B-V)_0^{\rm max}$&$M_B$&$(B-V)_0^{\rm max}$\\
\hline
\multicolumn{7}{c}{sequence re-calibrated in this work}\\
1965I  & 0.00 &31.30& $-18.89$& $-0.19$& $-18.80$&  $-0.23$\\
1970J  & 0.04 &33.20& $-18.24$& $+0.11$& $-18.90$&  $-0.26$\\
1971G  & 0.00 &31.73& $-17.83$& $-0.10$& $-17.53$&\phm0.00 \\
1972J  & 0.05 &32.83& $-18.12$& $+0.10$& $-18.53$&  $-0.28$\\
1975N  & 0.00 &31.81& $-17.81$& $+0.30$& $-18.21$&  $+0.17$\\
1975O  & 0.18 &33.80& $-18.68$& $+0.09$& $-18.50$&  $-0.23$\\
1976B$^{\#}$&0.00 &31.35& $-16.25$& $+1.24$& $-16.45$& $+1.17$\\
1976J  & 0.00 &33.52& $-19.24$& \phm0.00 & $-18.52$&  $-0.13$\\
1981B  & 0.00 &30.50& $-18.76$& $+0.06$& $-18.50$&  $-0.03$\\
\\
\multicolumn{7}{c}{sequence re-calibrated by Tsvetkov}\\
1969C  & 0.00 &32.99& $-19.20$& $+0.05$&  $-18.89$& $+0.12$ \\
1971L  & 0.24 &32.00& $-19.24$& $+0.19$&  $-18.90$& $+0.42$ \\
1972H  & 0.08 &32.79& $-18.47$& $+0.13$&  $-18.39$& $-0.23$ \\
1972R  & 0.06&30.12&$-17.33$&$+0.04$&$-16.82$&$-0.13$ \\
1973N  & 0.06 &33.77& $-18.92$& $-0.06$&  $-18.37$& $+0.37$ \\
1974G  & 0.00 &31.30& $-19.02$& $+0.30$&  $-18.80$& $ +0.17$\\
1974J  & 0.23 &34.30& $-18.93$& $-0.46$&  $-18.70$& $ -0.28$\\
1975G  & 0.00 &32.61& $-18.17$& $-0.15$&  $-17.81$& \phm0.00\\
1975P  & 0.00 &32.38& $-17.94$& $+0.26$&  $-17.78$& $+0.17$ \\
\\
\hline
\end{tabular}

\# Most likely a SN~Ib (Sandler \& Chugai 1986).\\
\end{center}\end{table*}

\begin{figure}
\psfig{figure=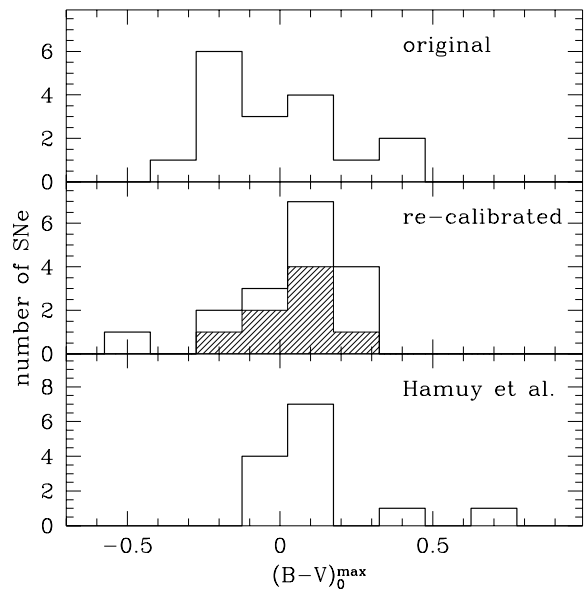,width=9cm}
\caption{\label{fig:col} Top panel: original
$(B-V)_0^{\rm max}$ color distribution before  
re-calibration. Middle panel: Re-calibrated color distribution. 
Shaded area indicates the SNe which sequences have
been re-calibrated in this work, the others are those of
Tsvetkov.  Bottom panel: $(B-V)_0^{\rm max}$ color distribution of 
the SN sample of Hamuy et al. (1995).}
\end{figure}

The effect of the color curve re-calibration on the distribution of the
$(B-V)_0^{\rm max}$ colors is illustrated in the upper two panels of
Fig.~\ref{fig:col}.  Excluding SN~1976B, which most likely is a SNIb
(Sandler \& Chugai 1986),
 the mean color $< (B-V)_0^{\rm max} >$ shifts from $-0.02$ to
 $+0.04$  and the dispersion decreases from 0.23 to 0.20 mag. The
re-calibrated distribution compares very well with that of the SN
sample by Hamuy et al. \cite{hamuy95} (Fig.~\ref{fig:col}, bottom
panel), which however has a redder average color:$<(B-V)_0^{\rm
max}>= +0.13$ and $\sigma(B-V)_0^{\rm max}=0.22$, due to the presence of
two red SN~Ia, 1990Y with $(B-V)_0^{\rm max}=+0.39$ and 1992K
with $(B-V)_0^{\rm max}=0.74$.

\begin{figure}
\psfig{figure=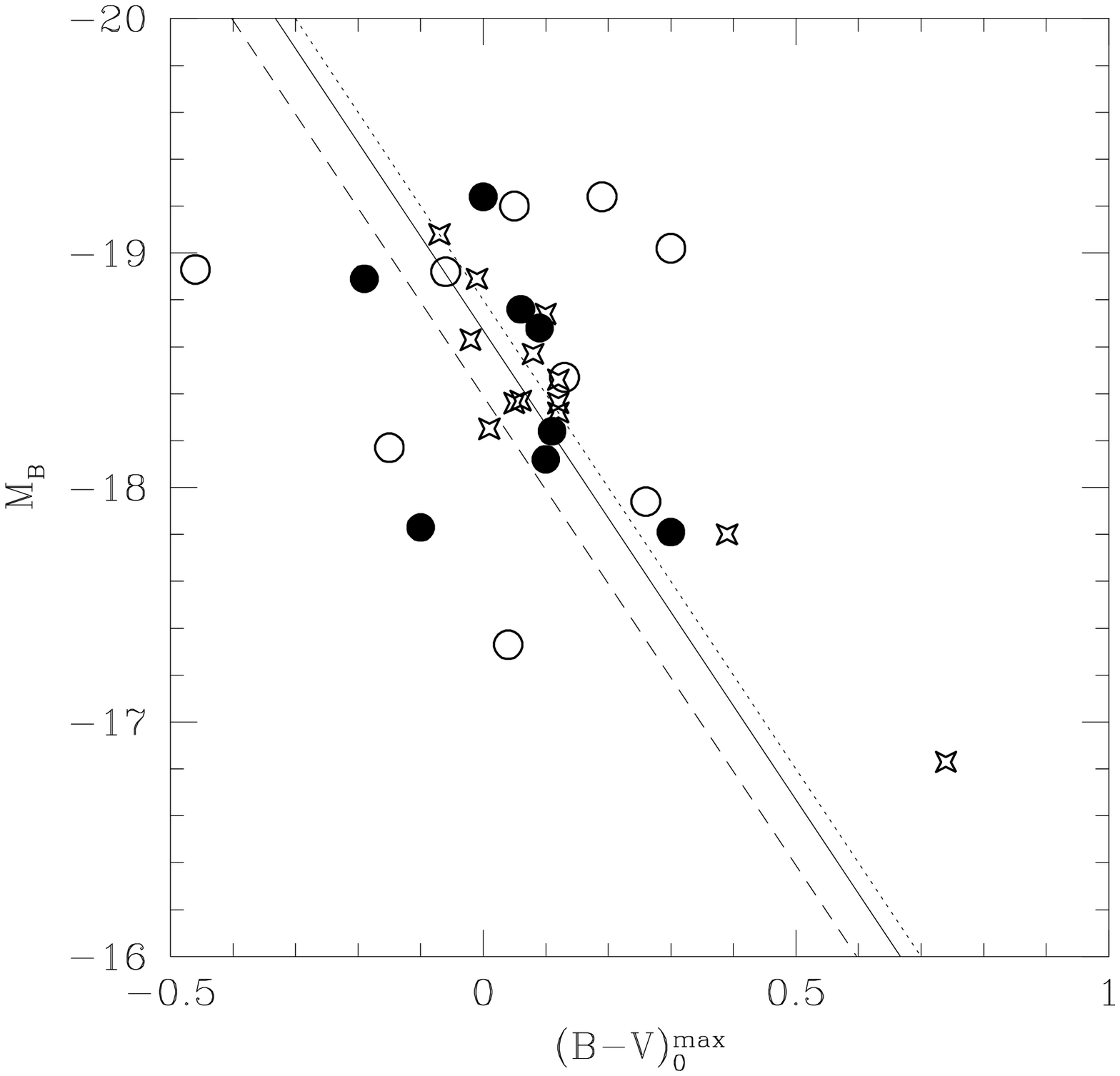,width=9cm}
\caption{\label{fig:final} $M_B$ vs. $(B-V)_0^{\rm max}$ for the Asiago SNe
whose sequences have been re-calibrated in this work (filled circles),
by Tsvetkov (empty circles) and by Hamuy et
al. (1995) (starred symbols). The continuous line is the fitting of 
the re-calibrated
Asiago data, the dashed line shows the fitting of the estimates based
on the original Asiago photometry (not shown) and the dotted line is
the fitting on the Hamuy et al.  sample.}
\end{figure}

We can now turn to the use of SN~Ia as
distance indicators. Let us assume that there are no intrinsic differences
among SN~Ia and that the observed differences are only due to photometric
errors and reddening.   We therefore expect that in the
 $M_B$ vs. $(B-V)_0^{\rm max}$ diagram the observed points distribute
around the line with slope $R = {A_B \over E(B-V)} = 4$, with a  dispersion
due to the photometric errors. This is shown in Fig.~\ref{fig:final}.

In order to derive the intrinsic luminosity of SN~Ia, assuming we know
the distances, we
need to identify a sub--sample of SNe for which we can measure, or
estimate, the reddening. This is not trivial, especially for small SN
samples affected by  significant photometric errors, and it explains
the large dispersion in the intrinsic color adopted by different
authors, from $-0.27$ (Cadonau et al. \cite{cadonau85}) to
$+0.09$(Sandage \& Tammann \cite{st93}).

On the other hand, if our aim is to use SN~Ia
as distance indicators, we could calibrate the $M_B$ vs. $(B-V)_0^{\rm
max}$ relation.  Since the slope of the relation is fixed, we
can refer for convenience to $M_B^{(B-V)=0}$, the absolute magnitude
corresponding to $(B-V)_0=0$, without knowing the intrinsic luminosity
and color of SN~Ia.  Then, in order to derive the distance modulus of
a new type Ia SN, we need only the difference between the observed
apparent magnitude at maximum and the expected absolute magnitude
corresponding to the observed SN color.
 The use of a larger SN sample in calibrating the $M_B$
vs. $(B-V)_0^{\rm max}$ relation strongly reduces the influence of
random photometric errors but, of course, it cannot help if 
 systematic errors are present.

For the re-calibrated SN sample of Asiago we compute $M_B^{(B-V)=0}=-18.67$.
This is in fair agreement with the value
$M_B^{(B-V)=0}=-18.80$ derived from the SN sample of
Hamuy et al. (1995) after excluding  SN~1992K which was recognized to be
intrinsically different from ``normal'' SN~Ia (Hamuy et al.1994).
The residual variance of the Asiago re-calibrated estimates and of the
data of Hamuy et al. (1995) around the regression lines drawn in
Fig.~6 is 0.21 and 0.07, respectively.

The difference of calibrations between the two SN samples corresponds
to only 6\% error in the distance calibration and the larger
dispersion of the Asiago re-calibrated sample has to be attributed to
the larger errors of the photographic photometry .  Instead, if we had used
the original Asiago data, we should have derived $M_B^{(B-V)=0}=-18.
39$ which, taking the data of Hamuy et al. (1995) as
reference, implies a 14\% error in the distance scale.

It should be stressed that a much larger error would result if, as
sometimes done in the past, the calibration is based
only on the bluest objects of the sample. In this case the result would 
be  dominated by photometric errors. 

The above considerations are valid in the hypothesis that SN~Ia have
intrinsically the same absolute magnitude and color at maximum. For
many years this assumption was made on the basis of both observational
and theoretical evidence. However, it is now accepted that SN~Ia are
not exactly alike. A few SN~Ia have shown strong photometric and
spectroscopic peculiarities, in particular the bright SN~1991T
(Ruiz-Lapuente et al. 1992, Phillips 1992, Filippenko et al. 1992a)  and
the faint SN 1991bg (Filippenko et al. 1992b, Leibundgut et al. 1993,
Turatto et al. 1996).  If these SNe are eliminated from the sample, the
dispersion of the SN absolute magnitude distribution strongly decreases but
the residual dispersion is not entirely due to photometric errors or
different reddening.  The two recent SNe 1992A and 1994D for which accurate
photometry is available, shows definite intrinsic differences (Patat et al.
\cite{nando}).  We also note that recently H\"oflich \& Khokhlov (1995)
have suggested, on the basis of theoretical modeling, the existence of a
relation between $M_B$ and $(B-V)_0^{\rm max}$ which mimics the reddening
law (van den Bergh \cite{van}). If this is the case, there is no need to
exclude extreme SN~Ia when calibrating the mentioned relation. 

Therefore, the intrinsic differences of SNIa
do not seem a severe threat to the use of SN~Ia as distance
indicators.
From the dispersion of the
points of the Hamuy et al.'s SN sample in the $M_B$ vs $(B-V)_0^{\rm max}$
diagram (which includes the photometric errors), it can be estimated
that the distance of a SNIa with accurate photometry can
be derived with a relative error of only 5\%.

\section{Conclusion}

In this paper we have presented and discussed new CCD observations of
several sequences of comparison stars used to calibrate the
photographic photometry of SNe observed in the past at the Asiago
Observatory. With these data, we have re-calibrated the light and color
curves of a sample of SN~Ia.  This work was motivated by the claims of
several authors (e.g. Sandage
\& Tammann \cite{st93}, Branch \& Miller \cite{bm93}, Vaughan
et al. (\cite{vaugh})) that systematic errors affected the photometry
of some SN~Ia observed at Asiago in the seventies.

Indeed we have found that, in most cases, the random errors affecting
the zero point calibration of the comparison stars were consistent
with the uncertainties quoted by the authors of the original papers
(cfr. Tables 4 and 5).  These errors ($0.1-0.2$ mag) were due to the
observing technique employed and which was based on photographic
transfers.  We have however  no explanation for the large errors found for the
sequence used to calibrate SNe 1970J and 1972J, whereas the sequence of
SN~1976J, the other object with a large correction, was reported as 
preliminary just in the original paper.

With the new sequences we have derived new light curves\footnote{The 
revised $B$ and $V$ SN photometry is
available upon request.} for the SNe
determining new $B$ magnitude and  $(B-V)$ color at maximum (see Tab. 6).
Large differences of these values from those ones of Leibundgut et al.
\cite{leib91} are found in several cases also because of different technique
of fitting.  As an effect of the re-calibration, the $(B-V)_0^{\rm
max}$ distribution becomes slightly sharper and moves to the red by
0.06 mag and it result in good agreement with that one of the SN sample of
Hamuy et al.(1995). Only one case, i.e. SN 1974J, of very blue supernova 
still remains which, as we said earlier, should need further 
investigation.

Finally, we stress that, as far as  the use of  SN~Ia as distance
indicators is concerned, the best approach  is to
calibrate the relation between $M_B$ and $(B-V)_0^{\rm max}$ using all
"bona fide" SN~Ia .  By using this approach, we cannot determine the
intrinsic absolute magnitude and color of the individual objects, but we
eliminate the problem of dealing with the unknown reddening in the parent
galaxy and strongly reduce the influence of random photometric errors.


\section*{Acknowledgments}

We are indebted to Dr. Dimitri Tsvetkov who kindly provided his data and
Prof. David Branch for stimulating discussions during the {\it NATO 
Advanced Study Institute on Thermonuclear Supernovae}.

\end{document}